\documentclass[aps,physrev,twocolumn,floats]{revtex4-2}
\usepackage{color,ulem,verbatim,float}
\usepackage{physics}
\usepackage{amssymb,amsbsy,amsmath,mathrsfs}
\usepackage{appendix}
\usepackage{feynmp-auto}
\usepackage{tikz}
\usepackage{tikz-feynman}
\tikzfeynmanset{compat=1.1.0}

\usepackage{graphicx} 

\usepackage{times}
\usepackage{color}
\usepackage{subfigure}
\usepackage{braket}
\usepackage{bbold}
\usepackage{commath}
\usepackage{setspace}
\usepackage{bm}

\def\red{\textcolor{red}}

\usepackage[colorlinks = true,
linkcolor = Blue,
urlcolor = Blue,
citecolor = Blue,
anchorcolor = Blue]{hyperref}
\usepackage{soul}
\usepackage[dvipsnames]{xcolor}

\newcommand{\dqmpgeneva}{Department of Quantum Matter Physics, University of Geneva, Quai Ernest-Ansermet 24, 1211 Geneva, Switzerland}

\begin{document}

\title{Interplay of Noise and Reservoir-induced Decoherence in Persistent Currents}

\begin{abstract}
Persistent current is a hallmark of quantum phase coherence. We study the fate of the persistent current in a non-equilibrium setting, where a tight-binding ring is subjected to stochastic disorder as well as a fermionic reservoir attached to each site. We evaluate the current using Keldysh technique and find that it exhibits non-monotonic behavior, suggesting two distinct mechanisms of decoherence. While coupling to the reservoirs introduces a coherence length scale given by the inverse of the coupling strength, the other mechanism is more subtle and driven by the ratio of noise strength to reservoir coupling. The interplay of noise and reservoir constitutes a purely non-equilibrium steady state with a flatter distribution function that we effectively describe using classical rate equations. We also discuss possibilities of realizing our findings in ultracold-atom experiments.
\end{abstract}

\author{Samudra Sur}
\affiliation{\dqmpgeneva}
\author{Thierry Giamarchi}
\affiliation{\dqmpgeneva}
\maketitle

\section{Introduction}
\label{intro}
Decoherence or loss of quantum coherence is ubiquitous as quantum many-body systems are never perfectly isolated. Interaction with the environment~\cite{feynman_vernon_2000,caldeira_leggett_brownian_1983,gardiner_quantum_noise_2004, clerk_introduction_rmp_2010}, presence of noise~\cite{gardiner_quantum_noise_2004,clerk_introduction_rmp_2010,medvedyeva_exact_prl_2016,bastianello_generalized_prb_2020}, and quantum measurements~\cite{skinner_measurement_prx_2019, alberton_entanglement_prl_2021, wiseman_quantummeasurement_2009,gurvitz_measurements_prb_1997, gurvitz_relaxation_prl_2003, carisch_quantifying_prr_2023,buchhold_effective_prx_2021, cao_entanglement_scipost_2019,zhang_universal_quant_2003} lead to loss of coherence, giving rise to emergent classical behavior. On the contrary, isolated disordered systems in equilibrium may undergo intrinsic decoherence due to electron-electron or electron-phonon interaction, resulting in suppression of interference effects like weak-localization~\cite{altshuler_phase_physicae_1998,gornyi_interacting_prl_2005,basko_metal_annal_2006}, and persistent current. 


Persistent current (PC) is one of the striking signatures of quantum phase coherence, arising when electronic wavefunctions remain coherent around a mesoscopic ring threaded by magnetic flux, producing the Aharonov–Bohm effect. Consequently, a suppression of the current indicates that one or several mechanisms of decoherence are at play. Although first introduced in early quantum mechanics~\cite{hund_persistent_adp_1938}, persistent currents gained significant attention only after their modern formulation in the context of mesoscopic systems~\cite{buttiker_josephson_pla_1983}. Over the last few decades, both analytical~\cite{cheung_isolated_ibm_1988,cheung_persistent_prb_1988,cheung_persistent_prl_1989,altshuler_persistent_prl_1991,schmid_persistent_prl_1991,vonoppen_average_prl_1991,ambegaokar_coherence_prl_1990,eckern_persistent_epl_1992, yu_persistent_prb_1992,fujimoto_persistent_prb_1993,smith_systematic_epl_1992,eckern_normal_advphys_1995,kopietz_universal_prl_1993,giamarchi_persistent_prb_1995,buttiker_small_prb_1985,trivedi_mesoscopic_prb_1988,wunsch_persistent_physicae_2004,semenov_persistent_prb_2009,patu_temperature_prl_2022} and numerical~\cite{bouchiat_persistent_jdp_1989, montambaux_persistent_prb_1990, bouchiat_persistent_prb_1991,bouzerar_persistent_prb_1995,schmitteckert_from_prl_1998} studies have elucidated the role of static impurities, interactions and temperature on the PC. Several experimental works~\cite{levy_magnetization_prl_1990,chandrasekhar_magnetic_prl_1991,jariwala_diamagnetic_prl_2001,deblock_diamagnetic_prl_2002,bluhm_persistent_prl_2009,bleszynski_persistent_science_2009} on isolated and ensembles of small metal rings have confirmed these understandings. Beyond fundamental interests, PC in ultracold atomic rings has also found applications~\cite{amico_colloquium_rmp_2022,polo_persistent_physrep_2025} in high-precision rotation sensing, atomtronic SQUID analogues (AQUIDs) for flux detection, and neutral persistent-current qubits.

While PC has been studied in the context of systems in thermodynamic equilibrium and in the presence of electron reservoirs~\cite{buttiker_small_prb_1985,trivedi_mesoscopic_prb_1988,wunsch_persistent_physicae_2004}, its behavior in non-equilibrium systems driven by the interplay of stochastic noise and reservoir coupling remains elusive. Such an interplay is known to give rise to non-equilibrium steady states~\cite{barontini_controlling_prl_2013, lueschen_signatures_prx_2017,tomita_observation_sciadv_2017,fitzpatrick_observation_prx_2017,ma_dissipatively_nature_2019,yamamoto_collective_prl_2021,dogra_dissipation_science_2019,ferri_emerging_prx_2021,syassen_strong_science_2008,sponselee_dynamics_qst_2018} marked by unusual transport~\cite{plenio_dephasing_njp_2008,rebentrost_environment_njp_2009,viciani_observation_prl_2015,maier_environment_prl_2019,corman_quantized_pra_2019,bredol_decoherence_prb_2021,huang_superfluid_prl_2023,visuri_symmetry_prl_2022,entin_effects_prb_2004,yamamoto_rectification_prr_2020, tonyjin_generic_prb_2020,tonyjin_exact_prr_2022,ferreira_transport_prl_2024,lacerda_transport_prb_2021}, dephasing~\cite{tonielli_orthogonality_prl_2019, mitchison_thermometry_prl_2020,dolgirev_dephasing_prb_2020,alba_unbounded_scipost_2022,wolff_evolution_prb_2019, lacerda_transport_prb_2021} and entanglement~\cite{skinner_measurement_prx_2019,alberton_entanglement_prl_2021,li_quantum_prb_2018,choi_quantum_prl_2020,gullans_dynamical_prx_2020,mueller_measurement_prl_2022,hoke_measurement_nature_2023,poboiko_theory_prx_2023} properties. In such systems PC can reveal the decoherence mechanisms due to the competing effects of noise and reservoir coupling. Moreover, considering the averaged dynamics, due to a correspondence between such stochastic systems and quantum systems under continuous weak measurements~\cite{caves_quantum_pra_1987,cao_entanglement_scipost_2019,bernard_transport_epl_2018,das_inverse_arxiv_2025} through a process known as \textit{unraveling}~\cite{dalibard_wavefunction_prl_1992,carmichael_open_1993,belavkin_nondemolition_1989,breuer_theory_2002}, PC can also probe the decoherence processes in weakly yet continuously monitored systems.

In this paper, we investigate the decoherence mechanisms that result in the decay of the persistent current in a ring of fermions subjected to stochastic white noise and fermionic reservoirs coupled to each site. Using a Keldysh formalism~\cite{kamenev_field_2011}, we find that noise and reservoir effects compete rather than add, giving rise to two distinct decoherence mechanisms exhibiting unusual decay of PC. Remarkably, the reservoirs play a dual role, with weak coupling supporting the current while strong coupling suppressing it. In the context of continuously monitored systems, the same mechanisms would prevail with noise strength replaced by the measurement rate.

The rest of the paper is organized as follows. In Sec.~\ref{model} we introduce the model of our study and also quantitatively define persistent current. In Sec.~\ref{Keldysh}, we introduce the Keldysh action to describe the non-equilibrium system, compute the effective contribution of the fermionic reservoirs, and also derive the effective action by performing the disorder averaging. In Sec.~\ref{scba} we discuss the self-energy induced by the stochastic noise and compute exact expressions for the full non-equilibrium Green functions of the system using the help of self-consistent Born method. In Sec.~\ref{persistent} we calculate the steady state persistent current and show that it shows non-monotonic behaviour due to two decoherence mechanisms. Then in Sec.~\ref{two mechanisms}, we describe the two decoherence mechanisms detail. Finally in Sec.~\ref{conclusion} we summarize and provide future perspectives. Some of the details of the calculations are provided in the Appendix.

\section{The model: Noisy Ring}
\label{model}

\begin{figure}[htb]
\centering
\includegraphics[width=0.48\textwidth]{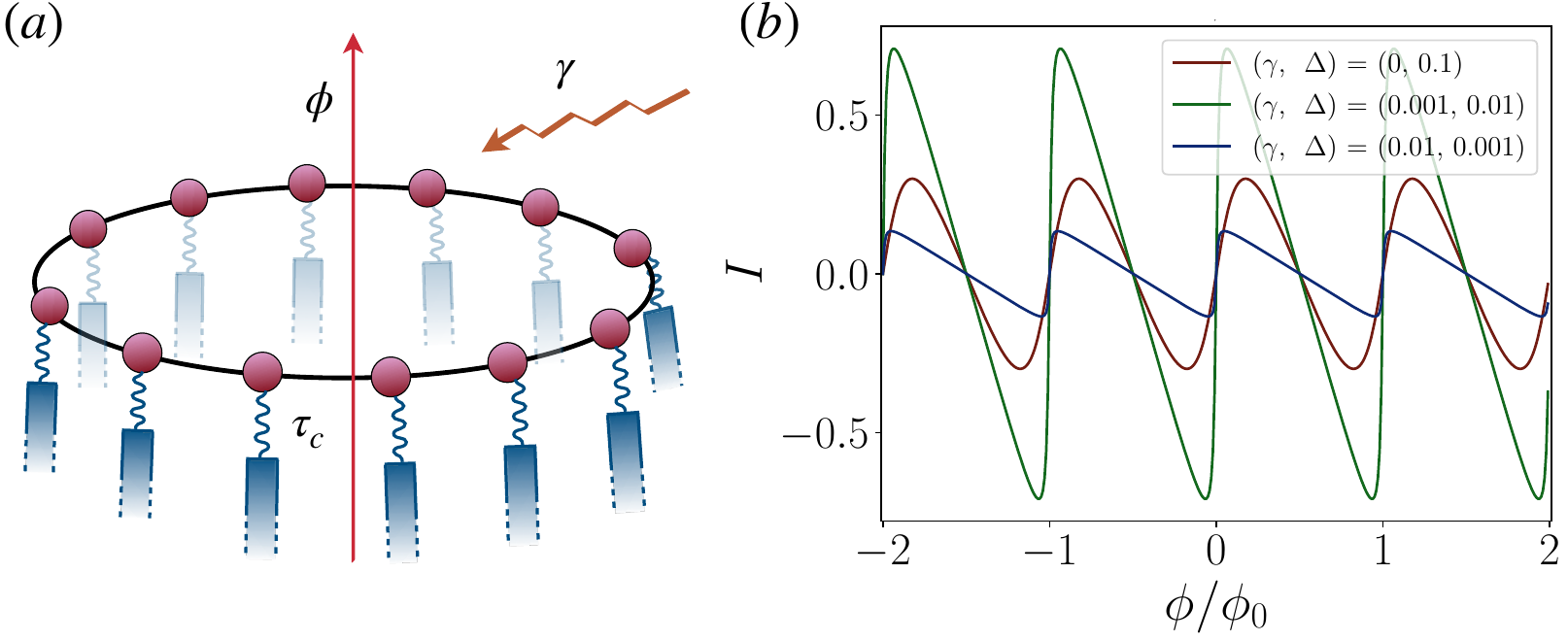}
\caption{\label{fig:schematic}(a) Schematic of the tight-binding ring in the presence of a time-dependent stochastic noise with strength $\gamma$ and identical one-dimensional fermionic reservoirs coupled to each site with coupling strength $\tau_c$. The ring is threaded by a magnetic flux $\phi$ that gives rise to the persistent current in the system. (b) Persistent current in units of $I_0 = ev_F/L$ as a function of flux $\phi/ \phi_0$ for three different sets of $(\gamma, \Delta) = (0,0.1), (0.001, 0.01), (0.01,0.001)$ in the units of $t$. $\Delta :$ bath hybridization function (see main text). The system size is $L=16$ with $N=8$ electrons.} 
\label{fig1} 
\end{figure}

We consider a tight-binding ring of spinless fermions on $L$ sites, subjected to a stochastic disorder that couples to the fermion density. Identical fermionic reservoirs with an energy dispersion $E_q$, chemical potential $\mu$, and temperature $T$ are connected to every site, which helps maintain a steady state in the presence of noise. \red{ }As the ring is threaded by a magnetic flux $\phi$, the Hamiltonian of the composite system is given by
\begin{align}
   \hspace{-6pt}&H(t) = -t\sum_{j=1}^{L}(e^{ieA}\psi^{\dagger}_{j}\psi^{\phantom{\dagger}}_{j+1} + \text{H.c.}) + \sum_{j=1}^{L}V_j(t)\psi^{\dagger}_{j}\psi^{\phantom{\dagger}}_{j} \nonumber \\
   \hspace{-4pt}+& \sum_{i=1}^{L}\sum_{q} (E_{q}-\mu) c^{\dagger}_{q,j} c^{\phantom{\dagger}}_{q,j} -\tau_c\sum_{j=1}^{L}\sum_{q}(\psi^{\dagger}_j c^{\phantom{\dagger}}_{q,j} +\text{H.c.}).
   \label{hamil}
\end{align}
The ring fermion ($\psi_j$) at site $j$ is coupled to the zeroth site of the corresponding reservoir ($c_{r=0,j} = \sum_{q}c_{q,j}$) with coupling strength $\tau_c$, while the tight-binding hopping in the ring is given by $t$. The magnetic flux is incorporated as a vector potential $A = \frac{2\pi}{L}(\phi/\phi_0)$, where $\phi_0$ is the flux quantum and $\hbar$ has been set to $1$ for convenience. The stochastic disorder $V_{j}(t)$ is a Gaussian white noise, with the noise correlation given by 
\begin{equation}
 \overline{V_i(t) V_j(t')} = 2\gamma ~\delta_{ij} \delta(t-t').  
 \label{noise_cor}
\end{equation}
A schematic of the model is shown in Fig.~\ref{fig1}(a) for illustration.
The total persistent current in the presence of magnetic flux is given as a weighted sum over the energy levels
$I(\phi) = \sum_n f(\mathcal{E}_n(\phi)) I_n(\phi),
$
where $f(\mathcal{E})$ is the distribution function and 
\begin{equation}
  I_n(\phi) = -\frac{e}{L} \frac{\partial \mathcal{E}_n(\phi)}{\partial k_n} = -e v_n(\phi)/L  
\end{equation}
is the current carried by $n^{\mathrm{th}}$ energy level~\cite{cheung_persistent_prb_1988} in the presence of flux. For a system in thermal equilibrium $f(\mathcal{E})$ would be given by the Fermi-Dirac distribution. Evidently, in a system with translational invariance, the net PC would be zero in the absence of magnetic flux, due to the exact cancellation of the currents carried by the pair of levels at $k$ and $-k$. However, the presence of a magnetic flux (non-integer multiple of $\phi_0$) breaks the time-reversal symmetry, thereby generating a non-zero PC. 

\section{Keldysh Action}
\label{Keldysh}

To study the non-equilibrium steady state due to the interplay of these two effects, we start with the Keldysh partition function~\cite{kamenev_field_2011} given by 
\begin{equation}
    Z = \int\mathcal{D}(\bar{\psi},\psi) \int \mathcal{D}(\bar{c},c) e^{iS(\bar{\psi},\psi,\bar{c},c)},
\end{equation} 
where the Keldysh action $S$ can be split into parts involving the system ($s$), bath ($b$) and coupling ($s\text{-}b$), i.e. 
\begin{equation}
  S = S_{s}(\bar{\psi},\psi) + S_{b}(\bar{c},c) + S_{s\text{-}b}(\bar{\psi},\psi,\bar{c},c).  
\end{equation} 
We assume that the Grassmann fields ($\bar{\psi}, \psi, \bar{c}, c $), defined on the upper and lower branches of the Keldysh contour have been rotated following Larkin-Ovchinnikov (LO) prescription~\cite{larkin_nonlinear_jetp_1975}. The quadratic nature of $S_b$ and $S_{s\text{-}b}$ allows us to integrate out the reservoir fermions and incorporate it as an effective term in $S_s$. In Fourier space, we perform the Gaussian integral involving the reservoir fermions, assuming a constant density of states (DOS) (the details of this derivation are provided in Appendix~\ref{appendix a}). This yields finally $S = S_s + S'(\bar{\psi}, \psi)$, where
\begin{align}  
\hspace{-5pt}S' =i \Delta \iint\frac{d\omega}{2\pi}\frac{dk}{2\pi} 
\begin{pmatrix}
   \bar{\psi}^1 & \bar{\psi}^2  
\end{pmatrix}
\begin{pmatrix}
 1 & 2\tanh[\frac{\omega-\mu}{2T}] \\
 0 & -1
\end{pmatrix}
\begin{pmatrix}
   \psi^1 \\
   \psi^2  
\end{pmatrix}
\label{bath_action}
\end{align}
The superscripts $1$ and $2$ denote the LO rotated Grassmann fields in Keldysh space, and we have omitted the $k,\omega$ labels for the fields for brevity. Furthermore, we introduce $\Delta = \tau_c^2/\mathbb{v}_F$, the bath hybridization function. Here $\mathbb{v}_F$, the Fermi velocity of the reservoir, is directly related to the constant DOS of the reservoir per unit length as $\rho_0 = 1/(\pi \mathbb{v}_F) $. The resulting full action can yet again be separated into two parts, $S = S_0 + S_{\mathrm{dis}}$, where $S_0$ governs the non-disordered part of the ring incorporating the effect of the reservoirs and $S_{\mathrm{dis}}$ describes the stochastic noise. The advantage of such separation is that the former, being time-independent and quadratic, can be expressed in terms of exact retarded, advanced and Keldysh Greens functions $g_{R/A/K}$. The entire procedure of integrating out the reservoirs amounts to the appearance of a finite lifetime and a temperature in the Greens functions of the ring fermions. More specifically, $S_0$ is given by
\begin{equation}
    S_0 = \int \frac{d\omega}{2\pi} \int \frac{dk}{2\pi} 
\begin{pmatrix}
   \bar{\psi}^1 & \bar{\psi}^2  
\end{pmatrix}
\begin{pmatrix}
    g_R^{-1} & [g^{-1}]_K \\
    0  & g_A^{-1}
\end{pmatrix}
\begin{pmatrix}
   \psi^1 \\
   \psi^2  
\end{pmatrix},
\label{clean_action}
\end{equation}
with $g^{-1}_{R(A)} = \omega -\mathcal{E}(k) \pm i\Delta$ and $[g^{-1}]_K = 2i\Delta \tanh(\frac{\omega-\mu}{2T})$. 

Heuristically, the bath hybridization $\Delta$ can now be understood as the effective imaginary self energy of an electron going from the ring to the reservoir and back. Naturally, this suggests that an electron in the ring has a finite lifetime given by the inverse of $\Delta$ due to the presence of reservoirs.

The disorder-averaging over the Gaussian distribution is performed at the level of the Keldysh partition function. In Fourier space, the probability distribution for the disorder potential can be expressed as $P[V] = \exp[-\frac{1}{4\gamma} \int \frac{dk}{2\pi}\int \frac{d\omega}{2\pi} ~V^*(k,\omega) V(k,\omega)]$, $V(k,\omega)$ being the Fourier transform of $V_i(t)$. The disorder averaged $$ \overline{Z}= \int\mathcal{D}(\bar{\psi},\psi) e^{iS_0} \int\mathcal{D}V P[V]e^{iS_{\mathrm{dis}}[V]} $$ gives rise to quartic terms in the action (see Appendix~\ref{appendix b} for the detailed derivation).
\begin{align}
  &\overline{Z} = \int\mathcal{D}(\bar{\psi},\psi)  e^{iS_0}  \exp \bigg\{ -\gamma \sum_{j} \int_{-\infty}^{\infty} dt \nonumber \\ &\big(\bar{\psi}^1_j(t)\psi^1_j(t) + \bar{\psi}^2_j(t)\psi^2_j(t)\big) 
  \big(\bar{\psi}^1_j(t)\psi^1_j(t) + \bar{\psi}^2_j(t)\psi^2_j(t)\big) \bigg\}
\end{align}
Note that, while the original action is not translationally invariant, disorder averaging with the given probability distribution restores the translational invariance. 

\section{Self-energy: Self Consistent Born Method}
\label{scba}

At the level of the Greens function, this effective \textit{interaction} arising from the disorder averaging, is incorporated through a self-energy  $\mathbf{\Sigma}$. Subsequently, the Dyson equation $\mathbf{G} = [\bm{\mathit{g}}^{-1}-\mathbf{\Sigma}]^{-1}$ relates the full Greens function with the bare Greens function $\bm{\mathit{g}}$ and self-energy $\mathbf{\Sigma}$.
The boldface indicates that it is a matrix equation with each quantity having $2\times2$ structure in Keldysh space. Diagrammatically, we represent the different components of $\bm{\mathit{g}}$ as follows,

$g_R =$
\begin{tikzpicture}
  \begin{feynman}
    \vertex [dot, minimum size=1.5pt] (i) {};
    \vertex [dot, right=1.4cm of i,minimum size=1.5pt] (o) {};
    \diagram* {
      (i) -- [anti fermion, arrow size=1.1pt,line width=0.7pt] (o),
    };
  \end{feynman}
\end{tikzpicture}
,\quad
$g_A =$
\begin{tikzpicture}
  \begin{feynman}
    \vertex [dot, minimum size=1.5pt] (i) {};
    \vertex [dot, right=1.4cm of i,minimum size=1.5pt] (o) {};
    \diagram* {
      (i) -- [anti charged scalar, arrow size=1.1pt,line width=0.7pt] (o),
    };
  \end{feynman}
\end{tikzpicture}
,\quad
$g_K = $
\begin{tikzpicture}
  \begin{feynman}
    \vertex [dot, minimum size=1.5pt] (i) {};
    \vertex [dot, right=0.7cm of i, minimum size=1.5pt] (v) {};
    \vertex [dot, right=0.7cm of v, minimum size=1.5pt] (o) {};
    \diagram* {
      (i) -- [black, thick] (v) -- [scalar, thick] (o),
    };
    \node at (v) [regular polygon, regular polygon sides=3, fill, rotate=-30, inner sep=1 pt] {};
  \end{feynman}
\end{tikzpicture}
    
In general, the self-energy $\mathbf{\Sigma}$ is given as a series of all one-particle irreducible diagrams at different orders. For instance, $\mathbf{\Sigma}_{R}$ has the following perturbative series expansion
\begin{equation}
\Sigma_{R} = 
\begin{tikzpicture}
  \begin{feynman}
    \vertex [dot, minimum size=1.5pt] (i) {};
    \vertex [dot, right=1cm of i,minimum size=1.5pt] (o) {};
    \diagram* {
      (i) -- [anti fermion, arrow size=1.1pt,line width=0.7pt] (o),
      (i) -- [photon, half left, looseness=1.3] (o)
    };
  \end{feynman}
\end{tikzpicture}
~+~
\begin{tikzpicture}
  \begin{feynman}
    \vertex [dot, minimum size=1.5pt] (i) {};
    \vertex [dot, right=0.5cm of i,minimum size=1.5pt] (v1) {};
    \vertex [dot, right=0.7cm of v1,minimum size=1.5pt] (v2) {};
    \vertex [dot, right=0.5cm of v2,minimum size=1.5pt] (o) {};
    \diagram* {
      (i) -- [anti fermion, arrow size=1.1pt,line width=0.7pt] (v1),
      (v1) -- [anti fermion, arrow size=1.1pt,line width=0.7pt] (v2),
      (v2) -- [anti fermion, arrow size=1.1pt,line width=0.7pt] (o),
      (i) -- [photon, half left, looseness=1.3] (o),
      (v1) -- [photon, half left, looseness=1.3] (v2)
    };
  \end{feynman}
\end{tikzpicture}
~+~
\begin{tikzpicture}
  \begin{feynman}
    \vertex [dot, minimum size=1.5pt] (i) {};
    \vertex [dot, right=0.5cm of i,minimum size=1.5pt] (v1) {};
    \vertex [dot, right=0.7cm of v1,minimum size=1.5pt] (v2) {};
    \vertex [dot, right=0.5cm of v2,minimum size=1.5pt] (o) {};
    \diagram* {
      (i) -- [anti fermion, arrow size=1.1pt,line width=0.7pt] (v1),
      (v1) -- [anti fermion, arrow size=1.1pt,line width=0.7pt] (v2),
      (v2) -- [anti fermion, arrow size=1.1pt,line width=0.7pt] (o),
      (i) -- [photon, half left, looseness=1.3] (v2),
      (v1) -- [photon, half left, looseness=1.3] (o)
    };
  \end{feynman}
\end{tikzpicture}
~+~ \cdots
\end{equation}
The wiggly lines represent the effective \textit{interaction} and the number of such lines is a measure of the order in the perturbation series. The series representation for $\Sigma_A$ can be obtained by replacing the solid straight lines with dashed lines. On the other hand, the Keldysh component $\Sigma_{K}$ has, in principle, more terms at each order depending on the position of the Keldysh component $g_K$.
\begin{equation}
\begin{aligned}
 \Sigma_K =~&
\begin{tikzpicture}
  \begin{feynman}
    \vertex [dot, minimum size=1.5pt] (i) {};
    \vertex [dot, right=0.5cm of i, minimum size=1.5pt] (v) {};
    \vertex [dot, right=0.5cm of v, minimum size=1.5pt] (o) {};
    \diagram* {
      (i) -- [black, thick] (v) -- [scalar, thick] (o),
      (i) -- [photon, half left, looseness=1.3] (o)
    };
    \node at (v) [regular polygon, regular polygon sides=3, fill, rotate=-30, inner sep=1 pt] {};
  \end{feynman}
\end{tikzpicture}
+
\begin{tikzpicture}
  \begin{feynman}
    \vertex [dot, minimum size=1.5pt] (i) {};
    \vertex [dot, right=0.4cm of i, minimum size=1.5pt] (v1) {};
    \vertex [dot, right=0.5cm of v1, minimum size=1.5pt] (v2) {};
    \vertex [dot, right=0.2cm of v2, minimum size=1.5pt] (v) {};
    \vertex [dot, right=0.3cm of v, minimum size=1.5pt] (o) {};
    \diagram* {
      (i) -- [anti fermion, arrow size=0.9pt,line width=0.7pt] (v1),
      (v1) -- [anti fermion, arrow size=0.9pt,line width=0.7pt] (v2),
      (v2) -- [black, thick] (v) -- [dashed, dash pattern=on 2pt off 1.5pt, thick] (o),      
      (i) -- [photon, half left, looseness=1.3] (o),
      (v1) -- [photon, half left, looseness=1.3] (v2)
    };
    \node at (v) [regular polygon, regular polygon sides=3, fill, rotate=-30, inner sep=0.8 pt] {};
  \end{feynman}
\end{tikzpicture}
+
\begin{tikzpicture}
  \begin{feynman}
    \vertex [dot, minimum size=1.5pt] (i) {};
    \vertex [dot, right=0.4cm of i, minimum size=1.5pt] (v1) {};
    \vertex [dot, right=0.3cm of v1, minimum size=1.5pt] (v) {};
    \vertex [dot, right=0.3cm of v, minimum size=1.5pt] (v2) {};
    \vertex [dot, right=0.4cm of v2, minimum size=1.5pt] (o) {};
    \diagram* {
      (i) -- [anti fermion, arrow size=0.9pt,line width=0.7pt] (v1),
      (v1) -- [black, thick] (v) -- [dashed, dash pattern=on 2pt off 1.5pt, thick] (v2),
      (v2) -- [dashed,dash pattern=on 2pt off 1.5pt, line width=0.7pt,postaction={decorate,decoration={markings,mark=at position 0.2 with {\arrowreversed[scale =0.9]{latex}}}}] (o),
      (i) -- [photon, half left, looseness=1.3] (o),
      (v1) -- [photon, half left, looseness=1.3] (v2)
    };
    \node at (v) [regular polygon, regular polygon sides=3, fill, rotate=-30, inner sep=0.8 pt] {};
  \end{feynman}
\end{tikzpicture}
+
\begin{tikzpicture}
  \begin{feynman}
    \vertex [dot, minimum size=1.5pt] (i) {};
    \vertex [dot, right=0.3cm of i, minimum size=1.5pt] (v) {};
    \vertex [dot, right=0.25cm of v, minimum size=1.5pt] (v1) {};
    \vertex [dot, right=0.5cm of v1, minimum size=1.5pt] (v2) {};
    \vertex [dot, right=0.4cm of v2, minimum size=1.5pt] (o) {};
    \diagram* {
      (i) -- [black, thick] (v) -- [dashed, dash pattern=on 2pt off 1.5pt, thick] (v1),
      (v1) -- [dashed,dash pattern=on 2pt off 1.5pt, line width=0.7pt,postaction={decorate,decoration={markings,mark=at position 0.2 with {\arrowreversed[scale =0.9]{latex}}}}] (v2),
      (v2) -- [dashed,dash pattern=on 2pt off 1.5pt, line width=0.7pt,postaction={decorate,decoration={markings,mark=at position 0.2 with {\arrowreversed[scale =0.9]{latex}}}}] (o),
      (i) -- [photon, half left, looseness=1.3] (o),
      (v1) -- [photon, half left, looseness=1.3] (v2)
    };
    \node at (v) [regular polygon, regular polygon sides=3, fill, rotate=-30, inner sep=0.8 pt] {};
  \end{feynman}
\end{tikzpicture} \\
&+
\begin{tikzpicture}
  \begin{feynman}
    \vertex [dot, minimum size=1.5pt] (i) {};
    \vertex [dot, right=0.4cm of i, minimum size=1.5pt] (v1) {};
    \vertex [dot, right=0.5cm of v1, minimum size=1.5pt] (v2) {};
    \vertex [dot, right=0.2cm of v2, minimum size=1.5pt] (v) {};
    \vertex [dot, right=0.3cm of v, minimum size=1.5pt] (o) {};
    \diagram* {
      (i) -- [anti fermion, arrow size=0.9pt,line width=0.7pt] (v1),
      (v1) -- [anti fermion, arrow size=0.9pt,line width=0.7pt] (v2),
      (v2) -- [black, thick] (v) -- [dashed, dash pattern=on 2pt off 1.5pt, thick] (o),      
      (i) -- [photon, half left, looseness=1.5] (v2),
      (v1) -- [photon, half left, looseness=1.3] (o)
    };
    \node at (v) [regular polygon, regular polygon sides=3, fill, rotate=-30, inner sep=0.8 pt] {};
  \end{feynman}
\end{tikzpicture}
+
\begin{tikzpicture}
  \begin{feynman}
    \vertex [dot, minimum size=1.5pt] (i) {};
    \vertex [dot, right=0.4cm of i, minimum size=1.5pt] (v1) {};
    \vertex [dot, right=0.3cm of v1, minimum size=1.5pt] (v) {};
    \vertex [dot, right=0.3cm of v, minimum size=1.5pt] (v2) {};
    \vertex [dot, right=0.4cm of v2, minimum size=1.5pt] (o) {};
    \diagram* {
      (i) -- [anti fermion, arrow size=0.9pt,line width=0.7pt] (v1),
      (v1) -- [black, thick] (v) -- [dashed, dash pattern=on 2pt off 1.5pt, thick] (v2),
      (v2) -- [dashed,dash pattern=on 2pt off 1.5pt, line width=0.7pt,postaction={decorate,decoration={markings,mark=at position 0.2 with {\arrowreversed[scale =0.9]{latex}}}}] (o),
      (i) -- [photon, half left, looseness=1.3] (v2),
      (v1) -- [photon, half left, looseness=1.3] (o)
    };
    \node at (v) [regular polygon, regular polygon sides=3, fill, rotate=-30, inner sep=0.8 pt] {};
  \end{feynman}
\end{tikzpicture}
+
\begin{tikzpicture}
  \begin{feynman}
    \vertex [dot, minimum size=1.5pt] (i) {};
    \vertex [dot, right=0.3cm of i, minimum size=1.5pt] (v) {};
    \vertex [dot, right=0.25cm of v, minimum size=1.5pt] (v1) {};
    \vertex [dot, right=0.5cm of v1, minimum size=1.5pt] (v2) {};
    \vertex [dot, right=0.4cm of v2, minimum size=1.5pt] (o) {};
    \diagram* {
      (i) -- [black, thick] (v) -- [dashed, dash pattern=on 2pt off 1.5pt, thick] (v1),
      (v1) -- [dashed,dash pattern=on 2pt off 1.5pt, line width=0.7pt,postaction={decorate,decoration={markings,mark=at position 0.2 with {\arrowreversed[scale =0.9]{latex}}}}] (v2),
      (v2) -- [dashed,dash pattern=on 2pt off 1.5pt, line width=0.7pt,postaction={decorate,decoration={markings,mark=at position 0.2 with {\arrowreversed[scale =0.9]{latex}}}}] (o),
      (i) -- [photon, half left, looseness=1.3] (v2),
      (v1) -- [photon, half left, looseness=1.5] (o)
    };
    \node at (v) [regular polygon, regular polygon sides=3, fill, rotate=-30, inner sep=0.8 pt] {};
  \end{feynman}
\end{tikzpicture}
+ \cdots
\end{aligned}
\end{equation}

Incidentally, the delta correlation of the disorder (Eq.~\eqref{noise_cor}) helps simplify the diagrammatic series for self-energy drastically, since all the diagrams involving a crossing of two wiggly lines vanish~\cite{tonyjin_exact_prr_2022}. This is a consequence of the causal (anti-causal) structure of the retarded (advanced) Greens function. Consequently, the self-energy consists of only the rainbow type diagrams, and can be exactly resummed using the Self-Consistent Born Approximation (SCBA)~\cite{dolgirev_dephasing_prb_2020,tonyjin_exact_prr_2022}, which is exact rather than an approximation in our case. Quantitatively, we can write in frequency space,
\begin{equation}
   \mathbf{\Sigma} = \gamma \int \frac{dk}{2\pi} \int \frac{d \omega}{2\pi} \mathbf{G}(k,\omega). 
\label{Sigma_SCBA}   
\end{equation}
In terms of diagrams, this implies,
\begin{equation}
\Sigma_R = 
\begin{tikzpicture}
  \begin{feynman}
    \vertex [dot, minimum size=1.9pt] (i) {};
    \vertex [dot, right=1.2cm of i,minimum size=1.9pt] (o) {};
    \diagram* {
      (o) -- [double,double distance=0.7pt, with arrow=0.5, arrow size=1.5pt,line width=0.7pt] (i),
      (o) -- [photon, half right, looseness=1.3] (i)
    };
  \end{feynman}
\end{tikzpicture}
~,
~~
\Sigma_A = 
\begin{tikzpicture}
  \begin{feynman}
    \vertex [dot, minimum size=1.9pt] (i) {};
    \vertex [dot, right=1.2cm of i,minimum size=1.9pt] (o) {};
    \diagram* {
      (o) -- [double,dashed, double distance=0.7pt, with arrow=0.5, arrow size=1.5pt,line width=0.7pt] (i),
      (o) -- [photon, half right, looseness=1.3] (i)
    };
  \end{feynman}
\end{tikzpicture}
~,
~~
\Sigma_K = 
\begin{tikzpicture}
  \begin{feynman}
    \vertex [dot, minimum size=1.5pt] (i) {};
    \vertex [dot, right=0.6cm of i, minimum size=1.9pt] (v) {};
    \vertex [dot, right=0.6cm of v, minimum size=1.9pt] (o) {};
    \diagram* {
      (i) -- [double, thick] (v) -- [double, dashed, thick] (o),
      (i) -- [photon, half left, looseness=1.3] (o)
    };
    \node at (v) [regular polygon, regular polygon sides=3, fill, rotate=-30, inner sep=1.3 pt] {};
  \end{feynman}
\end{tikzpicture}
\end{equation}
where the double lines indicate the full Greens function. We notice immediately from Eq.~\eqref{Sigma_SCBA} that the self-energy is not dependent on $\omega$ and $k$. This is again a signature of uncorrelated stochastic noise in space and time [Eq.~\eqref{noise_cor}]. The retarded and advanced components turn out to have very simple forms $\Sigma_{R(A)} = \mp i \gamma/2$, which is equal to the contribution from the first-order rainbow diagram. In fact, all the higher-order diagrams for $\Sigma_R$ and $\Sigma_A$ can be shown to be precisely equal to zero. The Keldysh self-energy, on the contrary, is slightly non-trivial to derive and depends on $T$ and $\mu$ of the reservoirs. Assuming $T=0$ for the reservoirs, we have (detailed calculations of the self-energies provided in Appendix~\ref{appendix c}) 
\begin{equation}
\Sigma_K = -\frac{4i \gamma}{\pi}  \int_0^{\pi}\frac{dk}{2\pi} \arctan\Big(\frac{\mathcal{E}(k)-\mu}{\Delta+\gamma/2}\Big) .
\end{equation}
The Dyson equation, along with the self-energy expressions, yields exact formulas for the different components of $\mathbf{G}$: 
\begin{align}
 G_{R(A)}(k,\omega) &= \frac{1}{\omega-\mathcal{E}(k)\pm i(\Delta + \gamma/2)}  \nonumber \\
 G_K(k,\omega) &= \lim_{T \to 0}\frac{\Sigma_K -2i\Delta \tanh(\frac{\omega-\mu}{2T}) }{[\omega -\mathcal{E}(k)]^2+(\Delta + \gamma/2)^2}
 \label{full_Greens}
\end{align}

\section{Steady state Persistent Current}
\label{persistent}

We now use these steady state Green functions to evaluate the persistent current.
Of particular interest is $G_K$, which determines (details provided in Appendix~\ref{appendix d}) the non-equilibrium steady state occupation function
$n(k) = \frac{1}{2}[1-i\int \frac{d\omega}{2\pi} G_K(k,\omega)]$.   

Instead of a discrete sum over the energy levels, we express PC as a fourier series in flux (details of the calculations to be found in Appendix~\ref{appendix e}) exploiting the flux periodicity.   
\begin{equation}
I(\phi) = -\frac{e}{2\pi} \sum_{m=-\infty}^{\infty} e^{i2\pi m\frac{\phi}{\phi_0}}\int^{\pi}_{-\pi} dk ~v(k)n(k) e^{-i(mL)k}
\label{persist2}
\end{equation}
where $v(k) = \partial{\mathcal{E}(k)}/\partial k$ is the electron velocity at $k$. Thus, the computation of PC requires the knowledge of the non-equilibrium distribution function. Evaluating \eqref{persist2} using $n(k)$ from \eqref{full_Greens}, we obtain the behavior of PC for different values of the noise strength $\gamma$ and bath hybridization $\Delta$. In Fig.~\ref{fig1}(b) we plot the PC (in units of $I_0 = ev_F/L$) as a function of flux $\phi/\phi_0$ for three sets of parameter values $(\gamma, \Delta) = (0,0.1), (10^{-3}, 10^{-2})$ and $(10^{-2}, 10^{-3})$, which shows typical oscillatory behavior with different amplitudes. We note that throughout the text, the numerical values of these parameters are taken to be in the units of $t$, the hopping parameter.

\begin{figure}[htb]
\centering
\includegraphics[width=0.48\textwidth]{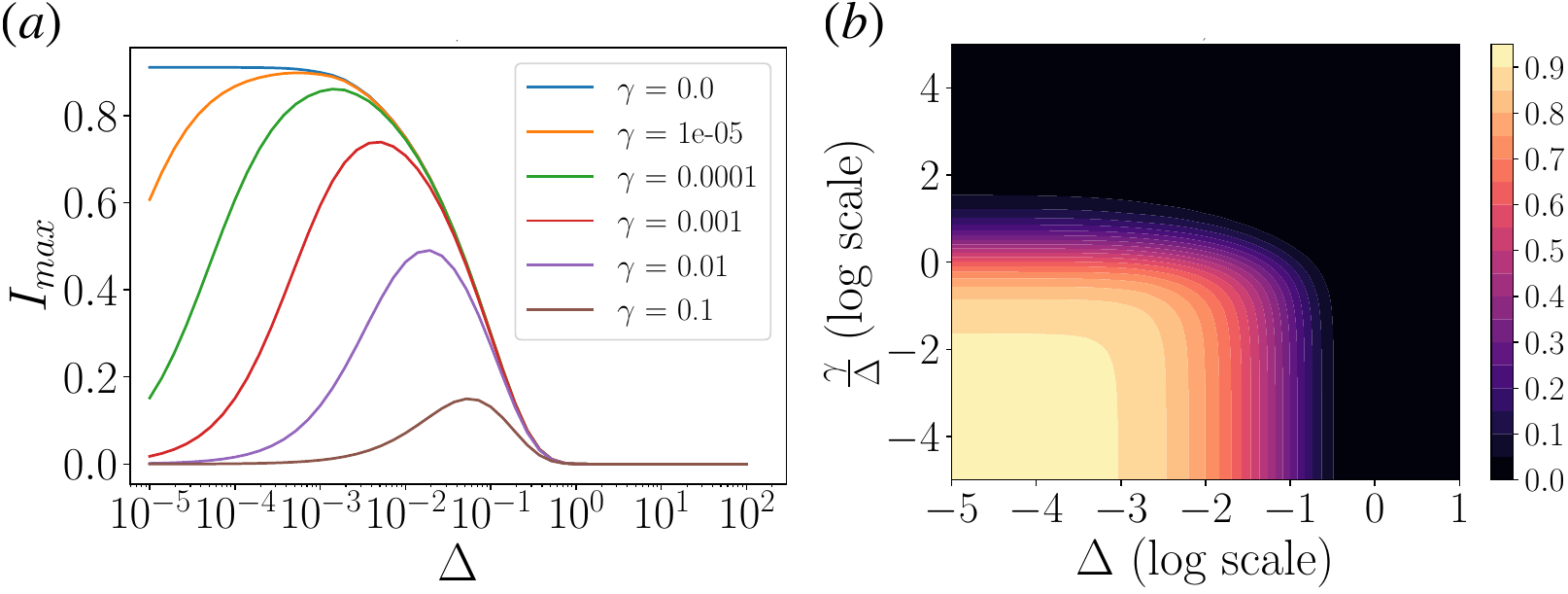}
\caption{(a) Amplitude of the persistent current ($I_{\mathrm{max}}$) plotted as a function of the bath hybridization function $\Delta$, for different values of the noise strength $\gamma$. For non zero $\gamma$, $I_{\mathrm{max}}$ exhibits non-monotonic behavior showing first a growth, then a peak, and finally decay with increasing $\Delta$. We fix $L=16$ and $N=8$, respectively. (b) Two independent and distinct mechanisms for the suppression of the persistent current are revealed, if $I_{\mathrm{max}}$ is plotted as a function of the pair of parameters ($\gamma/\Delta$, $\Delta$), as opposed to ($\gamma$, $\Delta$).} 
\label{fig2} 
\end{figure}

We now look at the variation of the amplitude of the PC ($I_{\mathrm{max}}$) at $T=0$, as a function of $\Delta$ and $\gamma$. Surprisingly, instead of an overall suppression from the additive effects of noise and the fermionic reservoir, they compete, leading to a unusual non-monotonic current. In Fig.~\ref{fig2}(a) we show $I_{\mathrm{max}}$ as a function of $\Delta$ in a semi-log plot for different noise strengths $\gamma = 0, 10^{-5}, \ldots ,10^{-1}$. Here $I_0$ is the current carried by an electron at the Fermi level, with $v_F$ being the Fermi velocity of the electrons in the ring. The system size is $L=16$ with $N = 8$ fermions. 

We observe in Fig.~\ref{fig2}(a) that for non-zero $\gamma$,  $I_{\mathrm{max}}$ first increases with increasing $\Delta$, reaches a maximum and then decays to zero. We also notice that the position of the peak of $I_{\mathrm{max}}$ shifts towards smaller $\Delta$ as $\gamma$ gets smaller and the maximum becomes flatter or plateau-like. Additionally, for $\gamma =0$, $I_{\mathrm{max}}$ shows monotonic behavior without any maximum at any $\Delta$. Furthermore, the decay at large $\Delta$ is seen to be almost independent of the value of $\gamma$. This suggests the existence of a sweet spot in the parameter space where, the PC is not appreciably affected. Nevertheless, two distinct pathways exist that destroy phase coherence in the system and suppress the PC.

\section{Two mechanisms of decoherence}
\label{two mechanisms}
The two independent routes for suppression of the PC become more apparent, once we look at the contour plot of $I_{\mathrm{max}}$ as a function of $\Delta$ and $\gamma/\Delta$, instead of $\gamma$ and $\Delta$, as shown in Fig.~\ref{fig2}(b). It clearly suggests that for a fixed $\Delta$, the ratio of noise to reservoir coupling $\gamma/\Delta$ provides an independent channel of decoherence and \textit{vice-versa}. In the absence of the reservoirs ($\Delta =0$), the presence of noise ($\gamma \neq 0$) will heat the system up to an infinite temperature steady state.   

\begin{figure}[htb]
\centering
\includegraphics[width=0.48\textwidth]{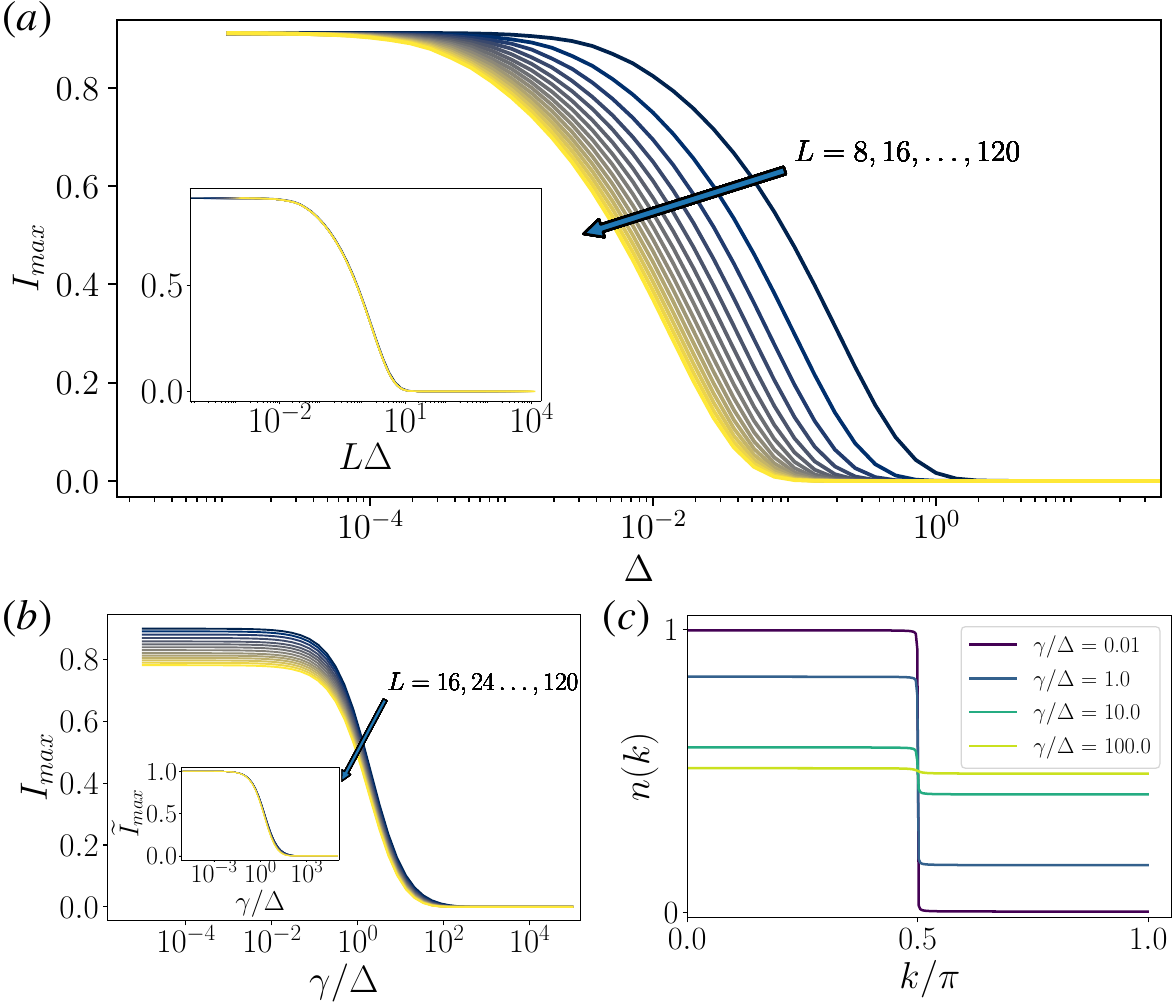}
\caption{(a) Plot for $I_{\mathrm{max}}$ as a function of $\Delta$ for different system sizes $L= 8, 16, \ldots,120$ at half-filling in the absence of noise ($\gamma= 0$). The curves show a systematic $L$ dependence and excellent scaling collapse if plotted as a function of $L\Delta$ [inset of (a)], leading to the notion of a coherence length scale $\xi_{\Delta} = v_F/\Delta$. (b) $I_{\mathrm{max}}$ as a function of the ratio $\gamma/\Delta$ for different $L = 16, 24,\ldots,120$ at half-filling for fixed $\Delta =0.001$. Except for the peak values, the curves are almost independent of $L$, which becomes clear if each of the curves is normalized by its peak value and plotted for different $L$[inset of (b)]. (c) The momentum occupation function $n(k)$ for different ratios $\gamma/\Delta = 0.01, 1, \ldots 100$ at a given $\Delta =0.001$. As $\gamma/ \Delta$ increases, $n(k)$ becomes progressively flatter, though the sharp jump at $k_F =\pi/2$ persists indicating the emergence of a pure non-equilibrium state distinct from a high-temperature equilibrium state.} 
\label{fig3} 
\end{figure}
We now turn to inspect each of these two mechanisms, driven by $\Delta$ and $\gamma/\Delta$ respectively, more closely. 

\subsection{Reservoir induced decoherence}

In Fig.~\ref{fig3}(a) we look at the behavior of $I_{\mathrm{max}}$ as a function of $\Delta$ for different system sizes $L = 8,16,\ldots 120$ at half-filling, in the absence of noise $\gamma$ or equivalently the ratio $\gamma/ \Delta = 0$. We immediately see that $I_{\mathrm{max}}$ shows a systematic system-size dependence, and the curves show a reasonably good scaling collapse once we rescale $\Delta$ by the system size $L$ [inset of Fig.~\ref{fig3}(a)]. We infer that for a fixed $\gamma/ \Delta$, the decay of maximum persistent current follows a universal function $I_{\mathrm{max}}/I_0 \sim \mathscr{I}(L\Delta) $, with $I_{\mathrm{max}}$ dropping to zero once $L\Delta$ exceeds $1$. This prompts us to define a coherence length scale $\xi_{\Delta} \sim 1/\Delta$ (modulo a factor of Fermi velocity $v_F \approx 2t$) induced by the reservoirs, beyond which the phase coherence of the system is lost and PC is suppressed. Equivalently, in terms of energy scales, reservoir induced decoherence becomes significant when $\Delta$ exceeds characteristic level spacing $\delta E= t/L$.

\subsection{Decoherence driven by noise to reservoir ratio}
In the regime of small reservoir coupling, the decoherence is driven by the ratio of noise to reservoir coupling $\gamma/\Delta$. The variation of $I_{\mathrm{max}}$ as a function of $\gamma/\Delta$, for fixed $\Delta = 10^{-3}$ shows little to no dependence on the system size $L$, except for the peak values, as shown in Fig.~\ref{fig3}(b).  Accordingly, the normalized $\widetilde{I}_{max}$, obtained by dividing each of the curves by the respective peak values, can be seen to be almost independent of $L$ as shown in the inset of Fig.~\ref{fig3}(b). Hence, this mechanism of decoherence cannot be associated with a length scale. However, the steady state occupation function $n(k)$ for different values of the ratio $\gamma/\Delta$ for given $\Delta = 10^{-3}$, exhibits interesting behavior. We notice in Fig.~\ref{fig3}(c) that as the ratio of noise to reservoir coupling $\gamma/ \Delta$ increases, states at $k >k_F = \pi/2$ start filling up, while states at $k \le k_F$ start getting depleted. This causes $n(k)$ to become flatter, although the discontinuous jump at the Fermi momentum $k_F$ remains. A flatter distribution implies that occupation probabilities are similar for states close in energy, carrying currents of either sign. The positive and negative contributions to the current cancel each other out significantly, leading to the suppression of the PC. In contrast, the distribution function of a higher temperature equilibrium state is rounded with a broadening of the edge at $k_F$. 

Empirically, this steady state can be understood using classical rate equations for momentum occupations. Spatially uncorrelated noise scatters fermions between filled and empty states regardless of momentum. As a result, the momentum distribution tends to be flatter. The retarded and advanced self-energies indicate that the scattering rate or strength is $\gamma/2$. Conversely, the reservoir will tend to relax the occupation of each momentum state towards the Fermi-Dirac distribution (at $T=0$) of the bath. Defining the average occupation above and below Fermi energy $\mu$ as $n_{>(<)} = \frac{2}{L}\sum_{|k| >(\le) k_F} n_k$, with the constraint $n_{>}+n_{<} =1$, we finally arrive at
\begin{align}
    \frac{d n_{<}}{dt} =& -\frac{\gamma}{4}(n_{<}-n_{>})- \Delta(n_{<}-1) \nonumber \\
    \frac{d n_{>}}{dt} =& -\frac{\gamma}{4}(n_{>}-n_{<})- \Delta ~n_{>}
\label{rate_eq}    
\end{align}

\begin{figure}[!htb]
\centering
\includegraphics[width=0.48\textwidth]{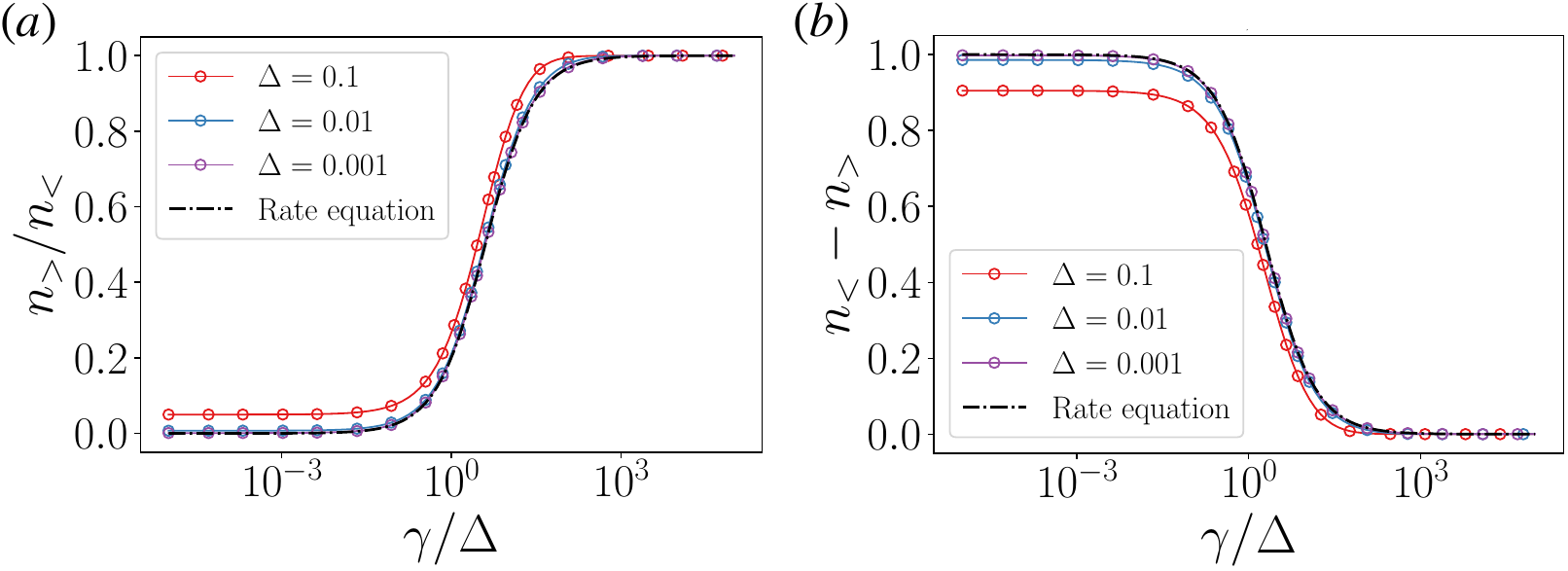}
\caption{(a) Ratio of steady-state average occupation above ($n_{>}$) and below ($n_{<}$) the Fermi energy obtained using Keldysh calculation for different $\Delta = 0.1, 0.01, ~\&~ 0.001 $ contrasted with the same obtained from the classical rate equations~\eqref{rate_eq}. (b) The difference of steady state average occupation $n_<-n_>$ for the same values of $\Delta$ as (a) contrasted with rate equation calculations. }
\label{fig4}
\end{figure}

In the steady state, the derivatives are zero as there are no transfer of particles. The average steady state population are then given by 
\begin{align}
  n_{<}= \frac{4\Delta+\gamma}{2(2\Delta+\gamma)}, ~~~~~ n_{>}= 1-n_{<}.  
\end{align}

In Fig.~\ref{fig4}, we contrast the ratio of occupations $n_{>}/n_{<}$ (a) and the difference of occupations $n_{>}$ and $n_{<}$ (b), obtained from rate equation calculations to those from Keldysh calculations for varied $\Delta = 0.1, 0.01, ~\&~ 0.001$. The two approaches show remarkable agreement with decreasing $\Delta$, suggesting that the classical rate equation can efficiently describe the emergent steady state due to the interplay of noise and reservoir.

\section{Conclusion and Perspectives}
\label{conclusion}

To summarize, the non-monotonic behaviour in the PC arises due to the competition between the stochastic noise and reservoir. In particular the origin of the two mechanisms can be understood from the distinct effects of the reservoir couplings. For weak reservoir coupling, the reservoirs essentially help maintain the Fermi-Dirac distribution of the ring electrons counteracting the noise that tends to flatten the distribution. Increasing $\Delta$ for a fixed noise $\gamma$, therefore initially enhances the current. At stronger coupling, however, the reservoirs broaden the energy levels and introduce a finite coherence length set by $\xi_\Delta =v_F/\Delta$. Once $\Delta$ becomes comparable to the characteristic level spacing $\delta E$ ($\sim t/L$), or equivalently, the coherence length scale becomes equivalent or shorter than the system size $\xi_\Delta \lesssim L$, reservoir-induced decoherence takes over. Therefore increasing $\Delta$ further supresses the persistent current irrespective of $\gamma$. Thus, the two regimes are controlled by the competing scales $\gamma/\Delta$ and $\Delta/\delta E$, respectively with the persistent current saturating in the window $\gamma \lesssim \Delta \lesssim \delta E$. 


A natural extension is to incorporate interactions in the non-equilibrium scenario, as in equilibrium, attractive interactions can enhance the current while repulsive interaction typically suppresses it~\cite{ambegaokar_coherence_prl_1990,giamarchi_persistent_prb_1995,semenov_persistent_prb_2009}. Another interesting direction would be to study the time evolution of the system towards this non-equilibrium steady state. Moreover, we believe our results could be accessible in ultracold-atom experiments, where realization of persistent currents in fermionic rings~\cite{pace_imprinting_prx_2022,polo_persistent_physrep_2025, amico_colloquium_rmp_2022}, fermionic reservoirs~\cite{brantut_conduction_science_2012,krinner_two_jpcmrev_2017}, and external noise~\cite{nagler_ultracold_prl_2022,hiebel_characterizing_njp2024} has already been possible.

\vspace{0.6cm}
\centerline{\bf Acknowledgments}
\vspace{0.2cm}
We thank Christophe Berthod for illuminating discussions and comments on the manuscript. We also acknowledge Géraldine Haack, Diptiman Sen and Catalin-Mihai Halati for useful discussions and suggestions. This work was supported by the Swiss National Science Foundation under Division II (Grant No. 200020-219400).

\appendix

\section{Derivation of effective action by integrating out the reservoir fermions}
\label{appendix a}

We present here the details of the derivation of the effective action by integrating the reservoir fermions. The Keldysh partition function of the system-reservoir composite is given as 
\begin{equation}
    Z = \int \mathcal{D}(\bar{\psi}, \psi) \int \mathcal{D}(\bar{c}, c) e^{i [S_{s}(\bar{\psi}, \psi)+ S_b(\bar{c},c) +S{s-b}(\bar{\psi}, \psi,\bar{c},c)]},
\end{equation}
where the partition function involving the bath and the system-bath coupling in Fourier space is given by 
\begin{align}
&Z_{bath} =  \int \mathcal{D}(\bar{c}, c)  e^{i (S_b+S_{s-b})} \nonumber \\
=& \int \mathcal{D}(\bar{c}, c) \exp \bigg[{i \sum_{j=1}^{L}} \int \frac{dq}{2\pi} \int \frac{d \omega}{2\pi}   
\begin{pmatrix} \bar{c}^{1}_{j}(q,\omega) & \bar{c}^{2}_{j}(q,\omega) \end{pmatrix} \nonumber \\
&~~~~~~~~~\begin{pmatrix}
   \omega - E_q +i \eta & 2 i \eta \tanh(\frac{\omega-\mu}{2T})  \nonumber \\
   0 & \omega - E_q -i \eta 
\end{pmatrix}
\begin{pmatrix}
    c^{1}_{j}(q,\omega)\\
    c^{2}_{j}(q,\omega)
\end{pmatrix} \\
& +  \tau_c \begin{pmatrix} \bar{c}^{1}_{j}(q,\omega) & \bar{c}^{2}_{j}(q,\omega) \end{pmatrix}
\begin{pmatrix}
    \psi^1_j(\omega)\\
    \psi^2_j(\omega)
\end{pmatrix} \nonumber \\
&
+\tau_c
\begin{pmatrix}
    \bar{\psi}^{1}_{j}(\omega) & \bar{\psi}^{2}_{j}(\omega)
\end{pmatrix}
\begin{pmatrix}
   c^{1}_{j}(q,\omega)\\
   c^{2}_{j}(q,\omega) 
\end{pmatrix}
\bigg].
\end{align}
We evaluate this Gaussian integral on the $(\bar{c},c)$ Grassmann fields using the Hubbard–Stratonovich transformation to obtain
\begin{align}
    Z_{bath} = \int &\mathcal{D}(\bar{c}, c) \exp \Bigg[ -i \tau_c^2\int \frac{d\omega}{2\pi} \sum_{j=1}^{L} \begin{pmatrix}
    \bar{\psi}^{1}_{j}(\omega) & \bar{\psi}^{2}_{j}(\omega)
\end{pmatrix} \nonumber \\
& \int \frac{dq}{2\pi}
\begin{pmatrix}
    G^R_{bath}(q,\omega) & G^K_{bath}(q,\omega)\\
    0 & G^A_{bath}(q,\omega)
\end{pmatrix}
\begin{pmatrix}
    \psi^1_j(\omega)\\
    \psi^2_j(\omega) 
\end{pmatrix}    
\Bigg].
\label{z-bath}
\end{align}
We introduce the reservoir Green functions $G^{R(A)}_{bath}(q,\omega) = \frac{1}{\omega - (E_q-\mu)\pm i\eta}$, and $G^K_{bath}(q,\omega) = -2\pi i \tanh(\frac{\omega-\mu}{2T}) \delta(\omega-E_q)$. Momentum integration on these Green functions requires the knowledge of the reservoir dispersion $E_q$. We assume that the reservoir dispersion to be linear with a corresponding Fermi momentum(velocity) given by $q_F (\mathbb{v}_F)$. The chemical potential is therefore given by $\mu = q_F$, since we have set $\hbar =1$. Consequently, the dispersion is given as
\begin{align}
    E_q = 
    \begin{cases}
        \mathbb{v}_F (q-q_F),& |q-q_F| \le \Lambda \\
        -\mathbb{v}_F (q+q_F), & |q +q_F| \le \Lambda,
    \end{cases}
\end{align}
where we have put a large cut-off $\Lambda$ in momentum for regularization. The integrals can now be performed with this dispersion,
\begin{align}
 \int \frac{dq}{2\pi} &G^{R(A)}_{bath}(q,\omega) = \int \frac{dq}{2\pi} \frac{1}{\omega -(E_q-\mu) \pm i\eta} \nonumber \\
 =& \int_{-\Lambda}^{\Lambda}\frac{dq}{2\pi} \frac{1}{\omega - \mathbb{v}_F q \pm i \eta } + \int_{-\Lambda}^{\Lambda}\frac{dq}{2\pi} \frac{1}{\omega + \mathbb{v}_F q \pm i \eta } 
\end{align}
Next, we take the limit $\Lambda \to \infty$ and use the Sokhotski–Plemelj theorem, which states  $\frac{1}{x+i \eta} = \text{P.V.}(\frac{1}{x}) \mp i\pi \delta(x) $ as $\eta \to 0$. Here P.V. denotes the Cauchy principle value. This yields finally
\begin{equation}
    \int \frac{dq}{2\pi} G^{R(A)}_{bath}(q,\omega) = \mp \frac{1}{\mathbb{v}_{F}}
\end{equation}
On the contrary, the Dirac delta function in $G^{K}_{bath}(q,\omega)$ can be easily integrated due to the linearity of dispersion
\begin{equation}
    \int \frac{dq}{2\pi} G^{K}_{bath}(q,\omega) = -\frac{2 i}{\mathbb{v}_F} \tanh{\Big(\frac{\omega-\mu}{2T}\Big)}
\end{equation}
The entire partition function can therefore be rewritten as $Z = \int \mathcal{D}(\bar{\psi}, \psi)e^{i(S_{s}+ S')}$, where $S'$ is the resulting action after integrating the reservoir fermions and has the following expression.
\begin{align}
    S' = & i\frac{\tau_c^2}{\mathbb{v}_F} \sum_{j=1}^{L}  \int \frac{d\omega}{2\pi} 
\nonumber \\
& \begin{pmatrix}
    \bar{\psi}^{1}_{j}(\omega) & \bar{\psi}^{2}_{j}(\omega)
\end{pmatrix} 
\begin{pmatrix}
    1 & 2\tanh(\frac{\omega-\mu}{2T})\\
    0 & 1
\end{pmatrix}
\begin{pmatrix}
    \psi^1_j(\omega)\\
    \psi^2_j(\omega) 
\end{pmatrix}. 
\end{align}
Subsequently, we introduce bath hybridization function $\Delta = \frac{\tau_{c}^2}{\mathbb{v}_F}$ and perform a Fourier transform of the fermionic fields as $\psi_j = \frac{1}{L} \sum_k \psi_k e^{-ikj}$, while the inverse Fourier transform is $\psi_k = \sum_{j=1}^{L} \psi_j e^{ikj}$. The Fourier transform of the $\bar{\psi}$ fields can be done by conjugation to ensure anticommutation relations in momentum space as well. Finally, taking the continuum limit in the discrete momentum sum $\sum_k \to L\int dk/2\pi$, yields Eqn.~(4) of the main text.

\section{Averaging the stochastic disorder: Effective Quartic interaction}
\label{appendix b}

The Keldysh action involving the stochastic disorder is 
\begin{align}
 S_{dis} &= -\sum_{j}\int_{\mathcal{C}} dt V_j(t) \bar{\psi}_j(t) {\psi}_j(t) \nonumber \\
 &= -\sum_{j} \int_{-\infty}^{\infty}dt V_j(t)[\bar{\psi}^1_j(t)\psi^1_j(t) + \bar{\psi}^2_j(t)\psi^2_j(t)] \nonumber \\
 &= -\sum_{j}\int_{-\infty}^{\infty} dt V_j(t) \rho_j(t),
\end{align}  
where $\mathcal{C}$ under the time integral denotes the Keldysh contour, the second equality is coming due to the LO rotation of the fields in the Keldysh space \cite{kamenev_field_2011}, and $\rho_j(t) = [\bar{\psi}^1_j(t)\psi^1_j(t) + \bar{\psi}^2_j(t)\psi^2_j(t)]$. The disorder $V_j(t)$ is a white noise that follows the Gaussian distribution
\begin{equation}
 P[V] = \exp \big(-\frac{1}{2} \sum_{i,j} \int_{-\infty}^{\infty} dt \int_{-\infty}^{\infty} dt' V_{i}(t) D^{-1}_{ij}(t-t')V_{j}(t) \big),   
\end{equation} 
with the variance of the distribution being $\overline{V_i(t)V_j(t')} = D_{ij}(t-t') = 2 \gamma \delta_{ij} \delta(t-t')$. Since in Keldysh formalism the observables are obtained from the Keldysh partition function as opposed to the logarithm of it, as in equilibrium formalism, the disorder averaging is performed over $Z$. If $S_0$ represents the action for the non-disordered part of system, the disorder averaging proceeds as follows,
\begin{align}
    \overline{Z} &= \int\mathcal{D}(\bar{\psi},\psi) e^{iS_0} \int\mathcal{D}V P[V]e^{iS_{dis}[V]}, \nonumber \\
=& \int\mathcal{D}(\bar{\psi},\psi) e^{iS_0} \int\mathcal{D}V \exp\bigg[ -i\sum_j\int_{-\infty}^{\infty} dt V_j(t) \rho_j(t) \nonumber \\
& -\frac{1}{2}\sum_{j,l} \int_{-\infty}^{\infty} dt' V_{j}(t) D^{-1}_{jl}(t-t')V_{l}(t) \bigg].
\end{align}
In the above expression, $\int \mathcal{D}V$ can be evaluated exactly, since it is a Gaussian integral, leading to an expression which is quadratic in $\rho$. This yields,
\begin{align}
   \overline{Z} =&  \int\mathcal{D}(\bar{\psi},\psi)  e^{iS_0} \exp\bigg[\frac{(-i)^2}{2} \nonumber \\
   &\sum_{l,j} \int_{-\infty}^{\infty} dt \int_{-\infty}^{\infty} dt' \rho_{j}(t) D_{jl}(t-t') \rho_l(t')  \bigg] \nonumber \\
   =& \int\mathcal{D}(\bar{\psi},\psi)  e^{iS_0} \exp\bigg[ -\gamma \sum_{j} \int_{-\infty}^{\infty} dt \rho_j(t) \rho_j(t) \bigg] \nonumber \\
  = & \int\mathcal{D}(\bar{\psi},\psi)  e^{iS_0} \nonumber \\
  &\exp\bigg[ -\gamma \sum_{j} \int_{-\infty}^{\infty} \big(\bar{\psi}^1_j(t)\psi^1_j(t) + \bar{\psi}^2_j(t)\psi^2_j(t)\big) \nonumber \\
  & ~~~~~~~~~~~~\big(\bar{\psi}^1_j(t)\psi^1_j(t) + \bar{\psi}^2_j(t)\psi^2_j(t)\big) dt \bigg]. 
\end{align}
We have used the delta function correlation of the disorder $D_{jl}(t-t')$.
Finally, Fourier transforming the fields we arrive at the effective action involving quartic interaction mentioned in the main text.

\section{SCBA self energy}
\label{appendix c}

We start with the bare Green functions given in Eq.~(5) of the main text, which denotes the non-disordered part of the action. As explained in the main text, SCBA provides an exact method to obtain the self energy in our case due to the delta correlation of the disorder (also see \cite{tonyjin_exact_prr_2022},\cite{dolgirev_dephasing_prb_2020}). 
The equation for SCBA self energy is 
$\mathbf{\Sigma} = \gamma \int \frac{dk}{2\pi} \int \frac{d \omega}{2\pi} \mathbf{G}(k,\omega)$,
which for retarded and advanced components becomes,
\begin{equation}
\Sigma_{R(A)} = \gamma \int_{-\pi}^{\pi} \frac{dk}{2\pi} \int_{-\infty}^{\infty} \frac{d\omega}{2\pi} \frac{1}{\omega -\mathcal{E}(k)\pm i\Delta -\Sigma_{R(A)}}. 
\end{equation}
The limits of the $\omega$ integral being $-\infty$ to $\infty$, the integral is fairly straightforward and is given by $\Sigma^{R(A)} = \mp \frac{i}{2} \gamma$. The full retarded and advanced Green functions becomes,
\begin{align}
    G_{R}(k,\omega) =& \frac{1}{\omega -\mathcal{E}(k)+ i (\Delta +\gamma/2)}, \nonumber \\
     G_{A}(k,\omega) =& \frac{1}{\omega -\mathcal{E}(k)- i (\Delta +\gamma/2)}.
\end{align}
The Keldysh self energy is slightly tricky to obtain. The SCBA equation for the Keldysh component of the self energy reads,
\begin{align}
    \Sigma_K =& \gamma \int_{-\pi}^{\pi} \frac{dp_1}{2\pi} \int_{-\infty}^{\infty} \frac{d\omega_1}{2\pi} G_K(p_1,\omega_1) \nonumber \\
    =&- \gamma \int_{-\pi}^{\pi} \frac{dp_1}{2\pi} \int_{-\infty}^{\infty} \frac{d\omega_1}{2\pi} \big[ G_R \big([g^{-1}]_K -\Sigma_K \big) G_A \big] \nonumber \\
    =& \gamma \int_{-\pi}^{\pi} \frac{dp_1}{2\pi} \int_{-\infty}^{\infty} \frac{d\omega_1}{2\pi} G_R(p_1,\omega_1) \Sigma_K G_A(p_1, \omega_1)  \nonumber \\
    &-\gamma \int_{-\pi}^{\pi} \frac{dp_1}{2\pi} \int_{-\infty}^{\infty} \frac{d\omega_1}{2\pi} \frac{2i \Delta \tanh(\frac{\omega_1 -\mu}{2T})}{(\omega_1 -\mathcal{E}(p_1))^2+(\Delta+\gamma/2)^2}.
\end{align}
Next, we take the limit of $T\to 0$, causing the $\tanh$ function to approach a step function asymptotically and therefore the integral in the second term becomes tractable. We also take the first term to the left hand side of the equation, and arrive at,
\begin{align}
  &\Sigma_K \bigg(1-\gamma \int_{-\pi}^{\pi}\frac{dp_1}{2\pi} \int_{-\infty}^{\infty} \frac{d\omega_1}{2\pi}\frac{1}{(\omega_1 -\mathcal{E}(p_1))^2+(\Delta+\gamma/2)^2}  \bigg) \nonumber \\
  &~~~~~~~= -2i\Delta\gamma \int_{-\pi}^{\pi}\frac{dp_1}{2\pi} \frac{2}{2\pi(\Delta+\gamma/2)}\arctan\bigg(\frac{\mathcal{E}(p_1)-\mu}{\Delta+\gamma/2}\bigg), \nonumber \\
& \implies \Sigma_K\bigg(1-\frac{\gamma}{2(\Delta+\gamma/2)}\bigg) \nonumber \\
&~~~~~~~~~~= -\frac{2i\Delta\gamma}{\pi(\Delta+\gamma/2)}\int_{-\pi}^{\pi}\frac{dp_1}{2\pi}\arctan\bigg(\frac{\mathcal{E}(p_1)-\mu}{\Delta+\gamma/2}\bigg).
\end{align}
After final rearrangements, this yields the expression for the Keldysh self energy at $T=0$ given in Eq.~(10) of the main text.
\begin{equation}
    \Sigma_K = -\frac{4i \gamma}{\pi}  \int_0^{\pi}\frac{dk}{2\pi} \arctan\Big(\frac{\mathcal{E}(k)-\mu}{\Delta+\gamma/2}\Big)
\end{equation}
It should be noted that the self energies are constant and not functions of $p$ \& $\omega$, due to the delta function correlation of the stochastic noise.

\section{Non-equilibrium occupation function from the Keldysh Green function}
\label{appendix d}

In a non-equilibrium steady state, i.e. when the correlation functions of the system only dependends on the relative time; the Keldysh Green function is related to the retarded and advanced Green function as,
\begin{align}
    G^K(k,\omega) =& F(\omega) [G^R(k,\omega) - G^A(k,\omega)] \nonumber \\
    =& (1-2f(\omega))[G^R(k,\omega) - G^A(k,\omega)],
\end{align}
where $f(\omega)$ is the non-equilibrium distribution function. The spectral function $A(\omega)$ is also related to the retarded and advanced Green function as 
\begin{equation}
    A(k,\omega) = i[G^R(k,\omega) - G^A(k,\omega)]
    \label{Ak}
\end{equation}
In addition, the spectral function also obeys the following sum rules,
\begin{equation}
    \int \frac{d\omega}{2\pi} A(k,\omega) = 1, ~~~~~\& ~~~~\int \frac{d\omega}{2\pi} f(\omega) A(k,\omega) = n(k).
    \label{sumrule}
\end{equation}
The second equation in \eqref{sumrule} and equation \eqref{Ak} trivially lead to the relation between $G^K(k,\omega)$ and $n(k)$
\begin{equation}
    n(k) = \frac{1}{2}[1-i \int_{-\infty}^{\infty}d\omega/2\pi G^K(k,\omega)]
\end{equation}

\section{Expression for the Persistent current as a Fourier Series in magnetic flux}
\label{appendix e}

In this section, we will derive the general expression for the persistent current, presented in Eq.~(12) of the main text. This expresses the persistent current as a Fourier series in the magnetic flux and the Fourier coefficients contain the occupation distribution, which can account for both equilibrium or non-equilibrium cases. Our starting point is the discrete sum in Eq.~(3) of the main text.
\begin{equation}
    I = -\frac{e}{L}\sum_n v_n(\phi) f(\mathcal{E}_n(\phi)).
\end{equation}
Here $f(\mathcal{E})$ is the distribution function. For a tight-binding ring with $L$ number of sites, the energy levels are $\mathcal{E}_n(\phi) = -2t \cos\big[\frac{2\pi}{L}(n+\phi/\phi_0)\big] = \mathcal{E}(n+\phi/\phi_0)$, and the subsequently the velocities of those levels are $v_n(\phi) = 2t \sin\big[\frac{2\pi}{L}(n+\phi/\phi_0)\big] = v(n+\phi/\phi_0)$, and $n$ ranges from $-L/2$ to $L/2-1$ . 

Let us define two functions $\zeta(n+\phi/\phi_0) = v(n+\phi/\phi_0) f(\mathcal{E}(n+\phi/\phi_0))$, and $\xi(\phi/\phi_0) = \sum_{n=-L/2}^{L/2-1} \zeta(n+\phi/\phi_0)$. Then, 
\begin{align}
   \xi(\phi/\phi_0 +1)&= \sum_{n= -L/2}^{L/2-1} \zeta(n+1+\phi/\phi_0) \nonumber \\
   &= \sum_{n=-L/2+1}^{L/2} \zeta(n+\phi/\phi_0) \nonumber \\
   = \xi(\phi/\phi_0)& + \zeta(L/2+\phi/\phi_0) -\zeta(-L/2+\phi/\phi_0).
\label{periodicity1}   
\end{align}  
However, $\zeta(L/2+\phi/\phi_0) - \zeta(-L/2+\phi/\phi_0) = 0$, since $v(L/2+\phi/\phi_0) = v(-L/2+\phi/\phi_0) $ and $\mathcal{E}(L/2+\phi/\phi_0) = -\mathcal{E}(-L/2+\phi/\phi_0) $. These relations along with Eq.~\eqref{periodicity1} leads to the fact that, 
\begin{equation}
    \xi(\phi/\phi_0 +1) = \xi(\phi/\phi_0),
\end{equation}
or in other words $\xi(\phi/\phi_0) $ is a periodic function with period $1$. This suggests that $\xi(\phi/\phi_0)$ admits a Fourier series in $\phi/\phi_0$.
\begin{align}
    \xi(\phi/\phi_0) &= \sum_{m=-\infty}^{\infty} \widetilde{\xi}(m) e^{i2\pi m\phi/\phi_0}, \nonumber \\
    \widetilde{\xi}(m) &= \int_0^1 dy ~\xi(y)~e^{-i2\pi my}
\end{align}
Thus, using the relation between $\zeta$ and $\xi$, we can write
\begin{align}
    \widetilde{\xi}(m) =& \int_0^1 dy \sum_{n=-L/2}^{L/2} \zeta(n+y)~e^{-i2\pi my} \nonumber \\
    =& \int_{-L/2}^{L/2} dy ~ \zeta(y)~e^{-i2\pi my} \nonumber \\
    =& \int_{-\pi}^{\pi}\frac{L}{2\pi}dk ~\zeta (\frac{Lk}{2\pi})~e^{-imLk},
\end{align}
where we have defined a new variable $k = \frac{2\pi y}{L}$. This re-definition helps simplifying the expression as $\zeta(\frac{Lk}{2\pi}) = v(k) f(\mathcal{E}(k))$, with $\mathcal{E}(k) = -2t\cos(k)$ and $v(k) = 2t\sin(k)$. We can now simply replace $f(\mathcal{E}(k))$ with $n(k)$, the momentum space occupation function.

The expression for persistent current can now be expressed as,
\begin{align}
   I =& -\frac{e}{L} \sum_{n} v_n(\phi) f(\mathcal{E}_n(\phi)) = -\frac{e}{L}\xi(\phi/\phi_0) \nonumber \\
=& -\frac{e}{L}\sum_{m=-\infty}^{\infty} \bigg\{\frac{L}{2\pi} \int_{-\pi}^{\pi}dk v(k) n(k) e^{-imLk} \bigg\} e^{i2\pi m\phi/\phi_0}    \nonumber \\
=& -\frac{e}{2\pi} \sum_{m=-\infty}^{\infty} e^{i2\pi m\frac{\phi}{\phi_0}}\int_{-\pi}^{\pi}dk ~v(k)n(k) e^{-i(mL)k}.
\end{align}
This is precisely the expression for persistent current given in Eq.~(12) of the main text.

\bibliography{my_references}

\end{document}